\documentclass[10pt,a4paper]{article}

\usepackage[utf8]{inputenc}
\usepackage[T1]{fontenc}
\usepackage{lmodern}
\usepackage{amsmath,amssymb,bm}
\usepackage{graphicx}
\usepackage{cite}
\usepackage[margin=0.85in]{geometry}
\usepackage[colorlinks=true,linkcolor=blue,citecolor=blue,urlcolor=blue]{hyperref}
\usepackage{placeins}

\title{Non-Forward OPE Kernel versus Scalar Form Factor in Near-Threshold
$J/\psi$ Photoproduction}

\author{A.~I.~Syamtomov\\[2pt]
\small Bogolyubov Institute for Theoretical Physics,\\
\small National Academy of Sciences of Ukraine, Kiev, Ukraine\\[2pt]
\small \texttt{arkady.syamtomov@gmail.com}}

\date{}

\begin{document}

\maketitle

\begin{abstract}
Near-threshold $J/\psi$ photoproduction has been used to infer scalar gluonic structure from integrated cross-section data. We test whether such an inference is stable against the assumed non-forward continuation of the OPE kernel. With an exponential diffractive slope fixed, the threshold fit favors an additional scalar $G_S(t)$-like contribution. However, the residuals show a systematic $W$-dependent trend. When either the effective slope or the kernel shape is replaced by a physically motivated form-factor continuation, the trend disappears and the fitted scalar coefficient is driven to zero. Thus integrated $\sigma(W)$ data cannot determine the non-forward OPE kernel or isolate a scalar $G_S(t)$ contribution. Differential $d\sigma/dt$ data are required.
\end{abstract}

\section*{1.\ Motivation and setup}

Near-threshold exclusive $J/\psi$ photoproduction, $\gamma p\to J/\psi\,p$, probes non-forward gluonic matrix elements: at threshold the physical $t$-interval collapses to a single value, $t_{\rm th}\simeq-2.2\,\mathrm{GeV}^2$, so even the forward endpoint is far from $t=0$. The near-threshold amplitude is commonly modeled as containing two gluonic contributions: a twist-two OPE component fixed by gluon PDF moments~\cite{Voloshin:1979uv,Peskin:1979va,Kharzeev:1996sq,Kharzeev:1999vh}, corrected for finite target mass~\cite{Nachtmann:1973mr}, and a possible scalar component usually associated with the QCD trace-anomaly gravitational form factor $G_S(t)$~\cite{Ji:1994av,Kharzeev:2021qkd}. GlueX measurements~\cite{GlueX:2019mkq,GlueX:2023pev} have been interpreted as evidence for this scalar structure and, more ambitiously, as a determination of the proton mass radius~\cite{Kharzeev:2021qkd,Boussarie:2020vmu,Gryniuk:2016mpk,Duran:2022xag}.

Define the $S$-wave projection of the dipole scalar form factor $G_S(t)=M_N/(1-t/m_s^2)^2$~\cite{Kharzeev:2021qkd}:
\begin{equation}
  \bar{G}_S(s) \equiv \frac{1}{2}\int_{-1}^{1}dz\,G_S(t(s,z)).
  \label{eq:Gbar}
\end{equation}
The threshold amplitude is then
\begin{equation}
  F_0(s)=\frac{P_0^{\rm OPE}(s)+C_S\,\bar{G}_S(s)}
              {1-i\rho_{\psi N}(s)\bigl[K_0^{\rm OPE}(s)+C_S\,\bar{G}_S(s)\bigr]},
  \label{eq:amplitude}
\end{equation}
where $\rho_{\psi N}=2k_\psi/\sqrt{s}$ is the phase-space factor and $K_0^{\rm OPE}$ is the target-mass-corrected twist-two kernel. For the present diagnostic the same scalar coefficient $C_S$ is used in the production and elastic kernels; relaxing this would only increase the degeneracy. The K-matrix denominator imposes a single-channel elastic final-state phase in the threshold window; it does not describe photoproduction outside $W\leq4.55$\,GeV. The integrated cross section vanishes as $k_\psi\to0$ with correct threshold behavior (no $1/k_\psi^2$ divergence%
\footnote{Explicitly, $\sigma(s)=(k_\gamma/k_\psi)\cdot3\Gamma_{ee}(s-s_{\rm th})^2/[16\pi\alpha_{\rm em}M_\psi(s-M_N^2)^2]\cdot|F_0|^2$. In this reduced-amplitude convention the explicit threshold factor overcompensates the $1/k_\psi$ prefactor, so the physical cross section vanishes at threshold and no $1/k_\psi^2$ divergence is present.}).

The forward OPE/PDF input does not determine the non-forward $t$-dependence; $K_0^{\rm OPE}(s)$ in Eq.~\eqref{eq:amplitude} equals $K^{\rm OPE}(s,0)$ times the $S$-wave projection of the assumed non-forward profile $F_{\rm OPE}(t)$:
\begin{equation}
  K_0^{\rm OPE}(s) = K^{\rm OPE}(s,0)\cdot\frac{1}{2}\int_{-1}^{1}dz\,F_{\rm OPE}(t(s,z)).
  \label{eq:K0OPE}
\end{equation}
Near threshold this non-forward continuation is a genuine model input. We therefore compare two families of choices, both satisfying $F_{\rm OPE}(0)=1$:
\begin{align}
  F_{\rm exp}(t;b) &= e^{b\,t/2}, \quad b(W)\approx2.5\,\mathrm{GeV}^{-2}\;\text{(diffractive exponential)}, \label{eq:Fexp}\\
  F_{\rm dip}(t;\Lambda) &= \bigl[\Lambda^2/(\Lambda^2-t)\bigr]^2 \quad\text{(VMD-/counting-rule-inspired dipole)}. \label{eq:Fdip}
\end{align}
The dipole gives $F_{\rm dip}(t_{\rm th};\Lambda=2\,\mathrm{GeV})\approx0.41$ and $F_{\rm dip}(t_{\rm th};\Lambda=M_\psi)\approx0.66$, whereas the diffractive exponential gives $e^{b\,t_{\rm th}/2}\approx0.06$ — a factor of 7--11 harder. The stability of the fitted scalar coefficient under this replacement directly tests whether integrated $\sigma(W)$ can isolate a scalar $G_S(t)$ contribution.

\section*{2.\ Integrated cross-section diagnostic}

All tests use nine GlueX-2023 points in $W_{\rm th}\leq W\leq4.55$\,GeV ($W=4.60$\,GeV excluded), with normalization and scalar coupling calibrated on this restricted set. Five kernel variants are compared (Table~\ref{tab:chi2}):

\begin{description}
\item[OPE/TMC baseline.] The dispersive twist-two amplitude with no scalar term; the reference curve.
\item[Fixed-slope K-matrix fit (A).] Exponential $F_{\rm exp}$ with diffractive $b(W)$, adding $G_S(t)$ and K-matrix phase. The fit favors a nonzero scalar coefficient ($\Delta\chi^2\approx9.7$, $\sim$3.1$\sigma$ relative to the baseline), but $\chi^2/\mathrm{dof}\approx5$ and the residuals show a systematic sign-flip trend — undershooting at low $W$, overshooting at high $W$.
\item[Freed-slope control (A$'$).] Exponential $F_{\rm exp}$ with freed $b_0+b_1(W-W_{\rm th})$. Removes the sign-flip trend; fitted $C_S\to0$, $\chi^2/\mathrm{dof}=0.05$.
\item[Dipole $\Lambda=2$\,GeV (B1) and $\Lambda=M_\psi$ (B2).] Form-factor $F_{\rm dip}$ with fixed $\Lambda$. Both remove the sign-flip trend and drive fitted $C_S\to0$, without freeing any slope parameter.
\item[Profiled dipole (D).] Dipole with $\Lambda$ varied freely. Prefers $\Lambda_{\rm best}\approx3.3$\,GeV; fitted $C_S\to0$, $\chi^2/\mathrm{dof}=0.07$.
\end{description}

\begin{table}[!htbp]
\centering
\begin{tabular}{llcccl}
\hline
Var. & Test & $\chi^2$ & $\chi^2/\mathrm{dof}$ & $C_S$ & $\langle F_0^{\rm proj}\rangle$ \\
\hline
— & OPE/TMC baseline (no scalar) & 49.2 & 6.15 & — & 0.058 \\
A  & Exp. fixed $b(W)$, $+G_S$, K-matrix & 39.5 & 4.94 & 6.5 & 0.058 \\
A$'$ & Exp. freed $b_0$, $+G_S$, K-matrix & 0.4 & 0.05 & 0 & 0.678 \\
B1 & Dipole $\Lambda=2\,\mathrm{GeV}$, $+G_S$, K-matrix & 3.5 & 0.44 & 0 & 0.342 \\
B2 & Dipole $\Lambda=M_\psi$, $+G_S$, K-matrix & 0.6 & 0.08 & 0 & 0.575 \\
D  & Dipole $\Lambda_{\rm best}=3.3\,\mathrm{GeV}$, $+G_S$, K-matrix & 0.6 & 0.07 & 0 & 0.611 \\
\hline
\end{tabular}
\caption{Threshold physics tests ($W\leq4.55$\,GeV, $N=9$, $\mathrm{dof}=8$). The column $\langle F_0^{\rm proj}\rangle$ is the mean S-wave projected kernel value over the nine GlueX fit points [Eq.~\eqref{eq:K0OPE}]. Variant A (diffractive exponential, $\langle F_0^{\rm proj}\rangle=0.058$) shows a sign-flip residual trend and favors $C_S\neq0$. All other variants have harder kernels ($\langle F_0^{\rm proj}\rangle\geq0.34$), flat pulls, and $C_S\to0$ — without any free slope parameter in B1, B2, D.}
\label{tab:chi2}
\end{table}

\begin{figure}[!htbp]
\centering
\includegraphics[width=\textwidth]{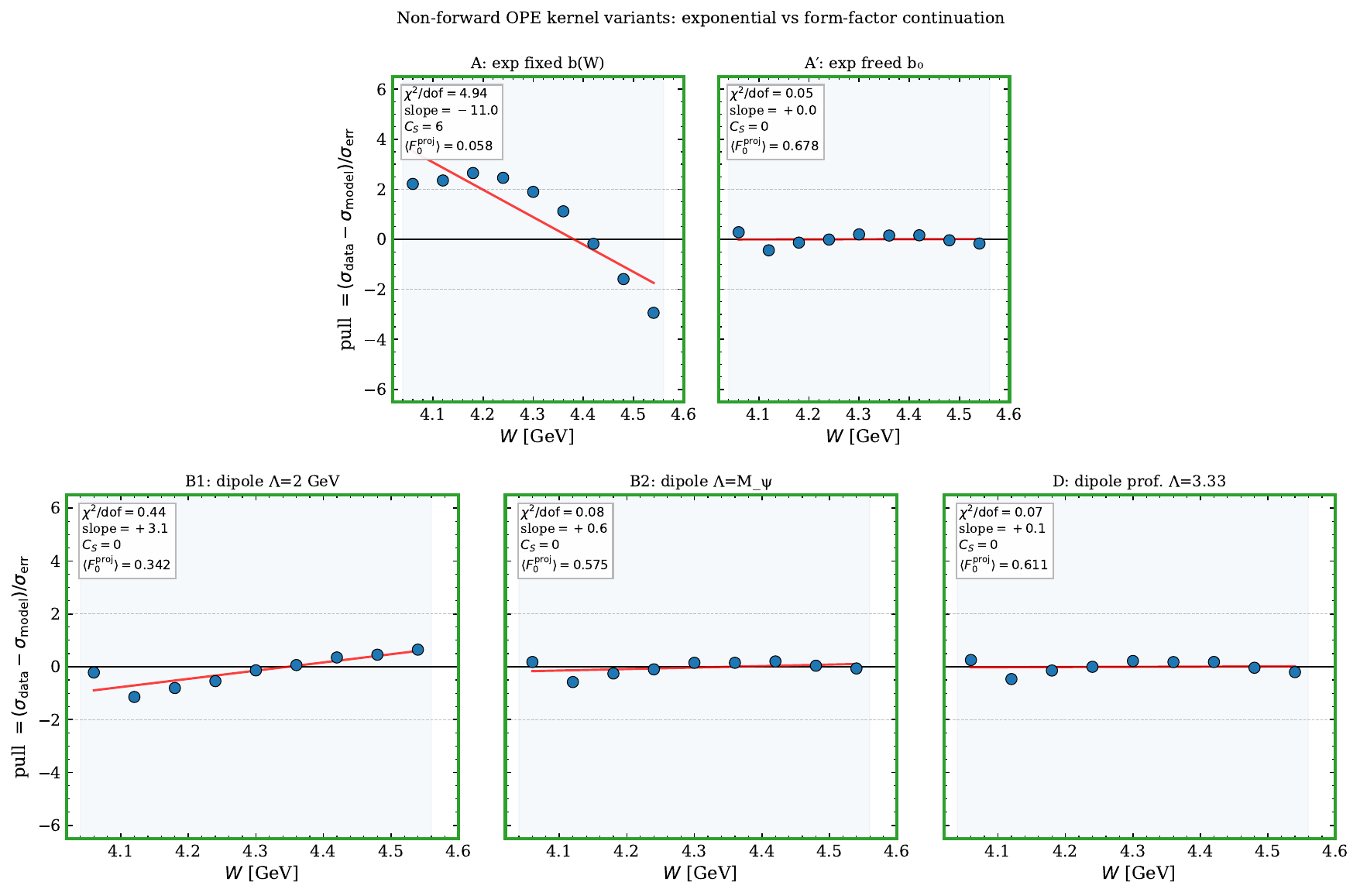}
\caption{Per-point pulls $(\sigma_{\rm data}-\sigma_{\rm model})/\sigma_{\rm err}$ for the five kernel variants. \textbf{A} (exp.\ fixed $b(W)$, $\langle F_0^{\rm proj}\rangle=0.058$): systematic sign-flip, fitted $C_S=6.5$. \textbf{A$'$} (exp.\ freed $b_0$, $\langle F_0^{\rm proj}\rangle=0.68$): flat, $C_S=0$. \textbf{B1} (dipole $\Lambda=2$\,GeV, $\langle F_0^{\rm proj}\rangle=0.34$): mildly sloped, $C_S=0$. \textbf{B2} (dipole $\Lambda=M_\psi$, $\langle F_0^{\rm proj}\rangle=0.58$): flat, $C_S=0$. \textbf{D} (dipole $\Lambda_{\rm best}=3.3$\,GeV, $\langle F_0^{\rm proj}\rangle=0.61$): flat, $C_S=0$. Green borders mark physically motivated variants. All variants except A drive the fitted $C_S$ to zero.}
\label{fig:kernel_pulls}
\end{figure}

\begin{figure}[!htbp]
\centering
\includegraphics[width=0.65\textwidth]{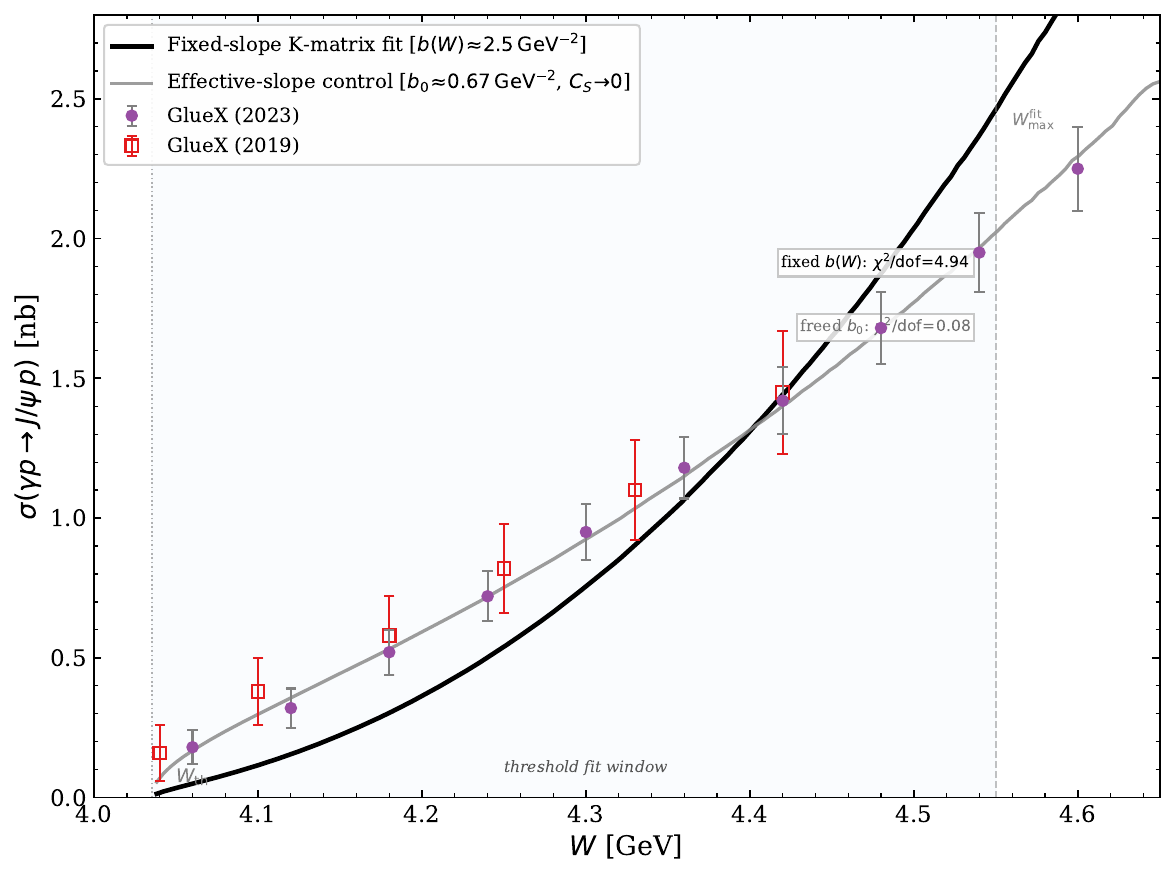}
\caption{Integrated $\sigma(\gamma p\to J/\psi\,p)$ vs.\ $W$. Threshold fit window ($W\leq4.55$\,GeV, shaded). Black solid: fixed-slope K-matrix fit (variant A, $\chi^2/\mathrm{dof}=4.94$). Grey solid: effective-slope control (variant A$'$, fitted $C_S\to0$, $\chi^2/\mathrm{dof}=0.05$). For clarity only the exponential fixed-slope and freed-slope controls are shown; the dipole-kernel variants (B1, B2, D) are compared in Fig.~\ref{fig:kernel_pulls} and Table~\ref{tab:chi2}. Data: GlueX-2023~\cite{GlueX:2023pev} (filled; $W=4.60$\,GeV excluded) and GlueX-2019~\cite{GlueX:2019mkq} (open).}
\label{fig:sigma_W}
\end{figure}

Figure~\ref{fig:kernel_pulls} shows the per-point pulls for all five kernel variants; Figure~\ref{fig:sigma_W} shows the integrated cross section. The root cause is visible in $\langle F_0^{\rm proj}\rangle$: the diffractive exponential kernel is very soft at threshold ($\approx0.06$), forcing the fit to compensate with $C_S\neq0$. Any harder non-forward continuation — whether a freed exponential slope or a form-factor dipole — describes the data without a scalar term.

The conclusion is direct: the apparent scalar preference in variant A is not a model-independent anomaly signal. It is induced by using a very soft non-forward OPE continuation at $|t|\sim2\,\mathrm{GeV}^2$, and is not stable under harder form-factor continuations. The degeneracy is broader than $b(W)$--$G_S(t)$: it is an underdetermination of the entire non-forward OPE kernel by integrated $\sigma(W)$. Note that $\Lambda_{\rm best}\approx3.3$\,GeV is an effective kernel-shape parameter inferred from integrated $\sigma(W)$, not a charmonium or gluonic radius. Additionally, the $m_s$ profile at fixed exponential kernel is flat for $m_s\gtrsim3$\,GeV, giving only $m_s\gtrsim3.26$\,GeV at $1\sigma$. No proton mass radius is extracted.

\section*{3.\ Implications}

\textbf{Integrated $\sigma(W)$ cannot determine the non-forward OPE kernel or isolate $G_S(t)$.} The forward OPE fixes the twist-two kernel only at $t=0$. Near threshold the physical point is already at $t_{\rm th}\simeq-2.2\,\mathrm{GeV}^2$, so the non-forward continuation is a genuine model input. Table~\ref{tab:chi2} shows that every continuation harder than the diffractive exponential ($\langle F_0^{\rm proj}\rangle\gtrsim0.3$) describes the data without a scalar term. The degeneracy is not merely $b(W)$--$G_S(t)$; it is an underdetermination of the whole non-forward OPE kernel. Neither the scalar coupling $C_S$ nor the mass scale $m_s$ can be extracted from integrated threshold cross-section data alone.

\begin{figure}[!htbp]
\centering
\includegraphics[width=0.78\textwidth]{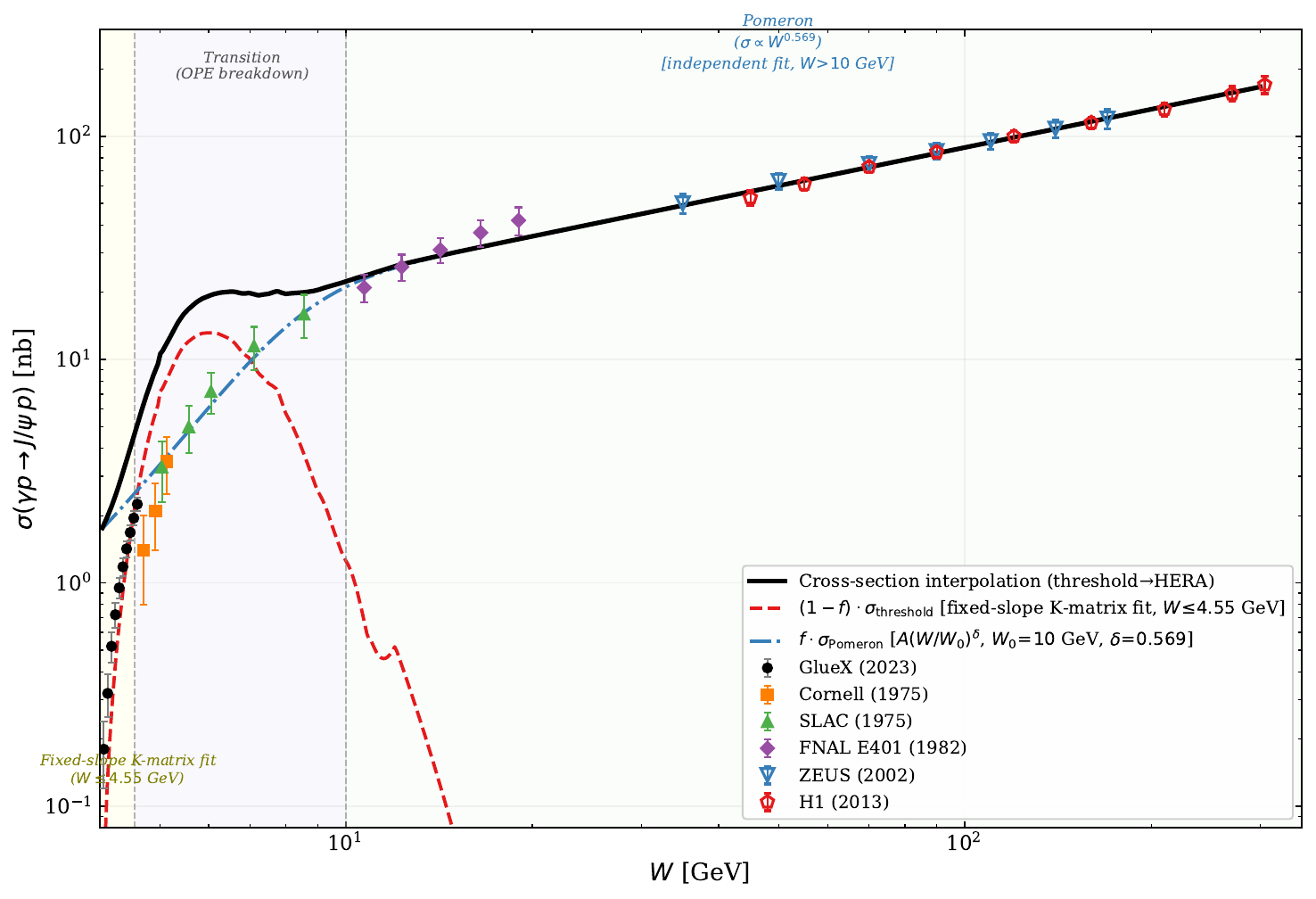}
\caption{Cross-section interpolation from threshold to HERA. Threshold ($W\leq4.55$\,GeV): fixed-slope K-matrix fit. High-energy ($W>10$\,GeV): independent cross-section fit $\sigma_P\propto W^{0.569}$ — no amplitude-level matching. Transition (4.55--10\,GeV): shown for continuity only. Data: GlueX~\cite{GlueX:2023pev}, Cornell~\cite{Gittelman:1975ix}, SLAC~\cite{Camerini:1975cy}, FNAL~E401~\cite{Binkley:1981kv}, ZEUS~\cite{ZEUS:2002wfj}, H1~\cite{H1:2013okq}.}
\label{fig:global}
\end{figure}

\textbf{High-energy data require separate treatment.} For $W>10$\,GeV an independent cross-section fit,
\begin{equation}
  \sigma_P(W)=A(W/W_0)^\delta,\quad W_0=10\,\text{GeV},\quad
  A=24.0\pm1.2\,\text{nb},\quad\delta=0.569\pm0.022,
  \label{eq:pomeron}
\end{equation}
($\chi^2/\text{ndf}=0.35$, 21 FNAL~E401+ZEUS+H1 points~\cite{Binkley:1981kv,ZEUS:2002wfj,H1:2013okq}), with no input from the OPE amplitude, describes the data well (Fig.~\ref{fig:global}). A cross-section interpolation $\sigma_{\rm interp}=(1-f)\sigma_{\rm th}+f\sigma_P$ connects the two regimes visually; the transition region is shown for continuity only. No coherent amplitude-level matching is claimed.

\textbf{Differential $d\sigma/dt$ is the required observable.} The non-forward kernel degeneracy is broken by the $t$-shape: the different interference patterns of $G_S(t)$ and $F_{\rm dip}(t;\Lambda)$ with the OPE term, together with their different running effective slopes $B_{\rm eff}(t)=d\ln(d\sigma/dt)/dt$, provide the discriminant. Measurements at $|t|\gtrsim1\,\mathrm{GeV}^2$ near threshold — where the large kinematic lever arm allows the two kernel shapes to separate — are therefore the key experimental test.

\textbf{Scope of the result.} The scalar form factor associated with the EMT trace is a well-defined QCD object. The point of the present analysis is more limited: integrated near-threshold $J/\psi$ photoproduction does not provide a clean way to isolate it. The observable is simultaneously sensitive to the scalar form factor and to the non-forward continuation of the twist-two OPE kernel. Since the latter is not fixed by the forward OPE/PDF input, different physically plausible kernel profiles describe the same integrated data while driving the fitted scalar coefficient to zero. Thus $J/\psi$ photoproduction remains a valuable gluonic probe, but integrated $\sigma(W)$ alone is not an ideal observable for extracting the scalar EMT form factor or a proton mass radius.

\clearpage

\end{document}